\documentclass[10pt]{article}

\usepackage{authblk}
\usepackage[square,numbers]{natbib}
\usepackage[T1]{fontenc}
\usepackage[utf8]{inputenc}
\usepackage{amsmath,amssymb,amsfonts}
\usepackage{algorithmic}
\usepackage{graphicx}
\usepackage[caption=false,font=footnotesize]{subfig}
\usepackage{textcomp}
\usepackage{url}
\usepackage{xurl}
\usepackage[htt]{hyphenat}
\usepackage{xcolor}
\usepackage{array}
\usepackage{booktabs}
\usepackage{braket}
\usepackage{enumitem}
\usepackage{environ}
\usepackage{ifthen}
\usepackage{makecell}
\usepackage{multirow}
\usepackage{pifont}
\usepackage{ragged2e}
\usepackage{tcolorbox}
\usepackage{xspace}
\usepackage{afterpage}
\usepackage[hidelinks]{hyperref}
\usepackage{geometry}
\usepackage{pdflscape}

\def\BibTeX{{\rm B\kern-.05em{\sc i\kern-.025em b}\kern-.08em
    T\kern-.1667em\lower.7ex\hbox{E}\kern-.125emX}}

\definecolor{light-gray}{gray}{0.80}

\newcommand{\Description}[1]{}
\newcommand{\markyes}{\ding{51}}
\newcommand{\markno}{\ding{55}}

\newcommand{\BenchmarkName}{{Qolumbina}\xspace}
\newcommand{\NumberOfPrograms}{40\xspace}
\newcommand{\NumberOfFamilies}{24\xspace}
\newcommand{\NumberOfRepositories}{8\xspace}
\newcommand{\Coverage}{85\%\xspace}
\newcommand{\DDL}{February 15, 2026}
\newcommand{\NumberOfUnitTests}{220\xspace}
\newcommand{\NumberOfProgramsInRQTwoTwo}{22\xspace}
\newcommand{\NumberOfProgramsInRQThreeOne}{23\xspace}
\newcommand{\NumberOfMixedTestsInRQThreeOne}{92\xspace}

\newcommand{\MaximumWidthOfRQThreeOne}{19\xspace}
\newcommand{\NumberOfProgramsInRQThreeTwo}{26\xspace}
\newcommand{\NumberOfCircuitsInRQThreeTwo}{1,680\xspace}
\newcommand{\MaximumWidthOfRQThreeTwo}{14\xspace}

\newcolumntype{C}[1]{>{\centering\arraybackslash}p{#1}}
\newcommand{\AnswerToRQ}[2]{%
  \begin{tcolorbox}[
    colback=black!1!white,
    colframe=black!40!white,
    left=0.25mm,
    right=0.25mm,
    top=0.25mm,
    bottom=0.25mm,
    boxsep=0.66mm,
    arc=0.1mm,
  ]
  \textbf{Findings of RQ#1:} #2
  \end{tcolorbox}
}

\newif\ifshowrevision
\showrevisionfalse
\newcommand{\revise}[1]{%
  \ifshowrevision
    \textcolor{blue!70!black}{#1}%
  \else
    #1%
  \fi
}

\NewEnviron{EnvRevise}{
  \ifshowrevision
    {\color{blue!70!black}\BODY}
  \else
    \BODY
  \fi
}

\newboolean{showcomments}
\setboolean{showcomments}{false}

\ifthenelse{\boolean{showcomments}}{
  \newcommand{\nbc}[3]{%
    {\colorbox{#3}{\bfseries\sffamily\scriptsize\textcolor{white}{#1}}}%
    {\textcolor{#3}{\sf\small$\langle$\textit{#2}$\rangle$}}%
  }
  
}{
  \newcommand{\nbc}[3]{}

}

\title{
Benchmarking Quantum Software Testing with Scalable Quantum Programs
}
\author[1,2,*]{Yuechen Li}
\author[2]{Minqi Shao}
\author[2]{Xiyuan Li}
\author[2]{Jianjun Zhao}
\author[1]{Kai-Yuan Cai}
\affil[1]{Beihang University\\
\texttt{\{liyuechen, kycai\}@buaa.edu.cn}}
\affil[2]{Kyushu University\\
\texttt{\{shao.minqi.229, li.xiyuan.868\}@s.kyushu-u.ac.jp}\\
\texttt{zhao@ait.kyushu-u.ac.jp}}
\affil[*]{Corresponding author}
\date{}

\begin{document}

\maketitle

\begin{abstract}
\revise{Quantum software testing (QST) checks whether quantum programs behave according to their intended specifications.
A key requirement for QST research is a benchmark that supports rigorous empirical evaluation on programs that are testable and better reflect current software development practices. 
However, existing studies heavily rely on small hard-coded or circuit-level benchmarks, while available quantum programs are scattered across repositories without clear selection criteria, which limits fair comparison and systematic reproducibility.
To this end, we present~\BenchmarkName, a benchmark infrastructure for controlled QST experiments on scalable quantum programs. 
\BenchmarkName curates~\NumberOfPrograms programs from open-source repositories, turns them into test-ready subjects through systematic selection, refactoring, specifications, test case examples, unit tests, and standardized interfaces.
We also propose QST-oriented criteria to characterize quantum programs along functionality, output behavior, development complexity, and quantum-specific execution complexity. 
Using these criteria, our empirical study shows that~\BenchmarkName covers diverse testing-relevant properties and supports scalability analysis beyond fixed-size circuit benchmarks.
Through controlled experiments with two recent QST approaches, we demonstrate the feasibility of using \BenchmarkName for execution-cost and fault-detection studies, and highlight backend-dependent effects that can influence QST result interpretation.}
\end{abstract}


\section{Introduction}\label{sec: intro}
With the advancement of quantum computing (QC), there is a growing need for high-quality quantum programs. 
Due to non-intuitive principles of quantum mechanics and increasing program scale, ensuring program quality and reliability remains challenging. As a solution in quantum software engineering (QSE)~\cite{zhao2020quantum, murillo2025quantum}, quantum software testing (QST)~\cite{10.1145/3377816.3381731} is a critical activity in the quantum software development life cycle (QSDLC), which assesses runtime behaviors of quantum programs and checks whether they satisfy given specifications.

Recently, empirical studies on QST have received growing attention. In such studies, benchmark quantum programs play a fundamental role in enabling fair and effective evaluation of testing approaches. 
Evaluating testing techniques requires benchmarks that go beyond small toy examples and cover programs with varying scales along with approximation to real-world scenarios.
However, many existing QST studies still rely on fixed-size quantum circuits (i.e., \textit{low-level quantum programs} in~\cite{li2026methodological}) as programs under test (PUT). These programs are often written in hardware-oriented assembly languages such as OpenQASM or expressed in low-abstraction code, making them poorly scalable and rarely maintainable in a practical QSDLC. 
{In contrast, scalable quantum programs (i.e., \textit{high-level quantum programs} in~\cite{li2026methodological}) employ richer abstractions and allow classical arguments to produce diverse quantum circuits.}
Their modular design more naturally follows software engineering (SE) practices, and such programs commonly appear in mainstream quantum software development kits (SDKs) such as Qiskit~\cite{Qiskit-qiskit}. Therefore, the gap between current research benchmarks and practical programs may limit the applicability of existing QST techniques in practical program-level testing scenarios.

{A recent study~\cite{li2026methodological} underscores that prior QST studies rely on programs collected from diverse and fragmented sources. Even the two most frequently used benchmarks are only partially aligned with QST. Bugs4Q~\cite{zhao2021bugs4q, zhao2023bugs4q} includes many classical programs in quantum software stacks, while MQT Bench~\cite{quetschlich2023mqt} was originally built for quantum hardware testing with quantum circuits as test inputs rather than PUTs. This purpose mismatch matters for software testing: when reused for QST experiments, such benchmarks provide limited support for functional specifications and standardized interfaces, making it difficult to design valid test cases and interpret test outcomes consistently.
Therefore, without a benchmark tailored to QST, it remains difficult to conduct comprehensive, fair, and reproducible evaluations across studies.}

{Motivated by these limitations, we propose a benchmark-construction methodology for controlled QST evaluation on scalable quantum programs, and instantiate it as~\BenchmarkName. 
It comprises~\NumberOfPrograms scalable Qiskit programs systematically filtered from~\NumberOfRepositories open-source repositories of programs with real-world provenance. 
To mitigate threats to testability and empirical validity arising from current quantum programming practices, including heterogeneous program interfaces and under-specified input constraints, the original programs are moderately refactored into standardized and executable benchmark subjects by improving interface compatibility, aligning language usage, and validating test inputs. 
To enhance the functional testability, we provide program specifications that document intended program behavior, together with unit tests as executable examples.}

{We further conduct a systematic empirical study to characterize the involved quantum programs and examine testing-relevant properties of scalable quantum programs.}
{From the perspective of QST, we establish a set of criteria for analyzing program functionalities in terms of application domains and output characteristics.} Empirical evidence reveals the functional diversity of~\BenchmarkName based on the covered categories and indicates that the design of testing approaches should take such differences into account, although this issue has received limited attention in prior QST research. 
Then, we characterize program scale along both development and execution dimensions, using standard SE metrics overlooked in QST (e.g., lines of code (LOC)) and quantum-specific metrics (e.g., circuit width). The results show that, unlike fixed-size subjects, our benchmark programs instantiate circuits whose scale depends on classical inputs.
{Finally, we conduct two controlled experiments in both noiseless and noisy scenarios to examine the feasibility of using \BenchmarkName in QST experiments. 
By applying two recent QST approaches~\cite{li2025preparation,li2016dynamic} and aligning with their scopes, our experiments consider \NumberOfProgramsInRQThreeOne programs with sufficient test suites for execution-cost analysis and \NumberOfProgramsInRQThreeTwo programs with 7 buggy variants each for fault-detection analysis.
Experimental results demonstrate that \BenchmarkName can reproduce findings from prior studies and also provide insights into how fake backends affect QST result interpretation.
}

{The main contributions of this paper are listed as follows:}

\begin{itemize}[leftmargin=*]
  \item We present~\BenchmarkName, a publicly available benchmark infrastructure for controlled QST research on scalable quantum programs. It contains~\NumberOfPrograms scalable Qiskit programs that are closer to practical quantum software development than the low-abstraction and circuit-like programs commonly used in prior studies.
    
  \item {Instead of simply collecting programs, we propose a construction pipeline that turns open-source quantum programs into test-ready benchmark subjects through systematic selection, program refactoring, program specifications, unit tests, and standardized interfaces.}
    
  \item {We establish systematic QST-oriented taxonomies and criteria for characterizing program functionalities and scales, and also use them to validate benchmark diversity and scalability support of~\BenchmarkName.}

  \item {We conduct two controlled experiments with recent QST approaches, demonstrating the feasibility of using~\BenchmarkName and providing empirical findings for future QST research.}
\end{itemize}

The remainder of this paper is organized as below. Section~\ref{sec: background} introduces preliminaries of quantum programs and reviews related benchmarks. Section~\ref{sec: methodology} presents the methodology for infrastructure construction. Section~\ref{sec: empirical} reports the empirical study, and Section~\ref{sec: threats} discusses threats to validity. Section~\ref{sec: discussions} outlines lessons for future research. Finally, Section~\ref{sec: conclusion} concludes the paper.

\section{Background and Related Work}\label{sec: background}
\subsection{Quantum Programs and Their Design}
\begin{figure}[htbp]
    \centering
    \includegraphics[width=\columnwidth]{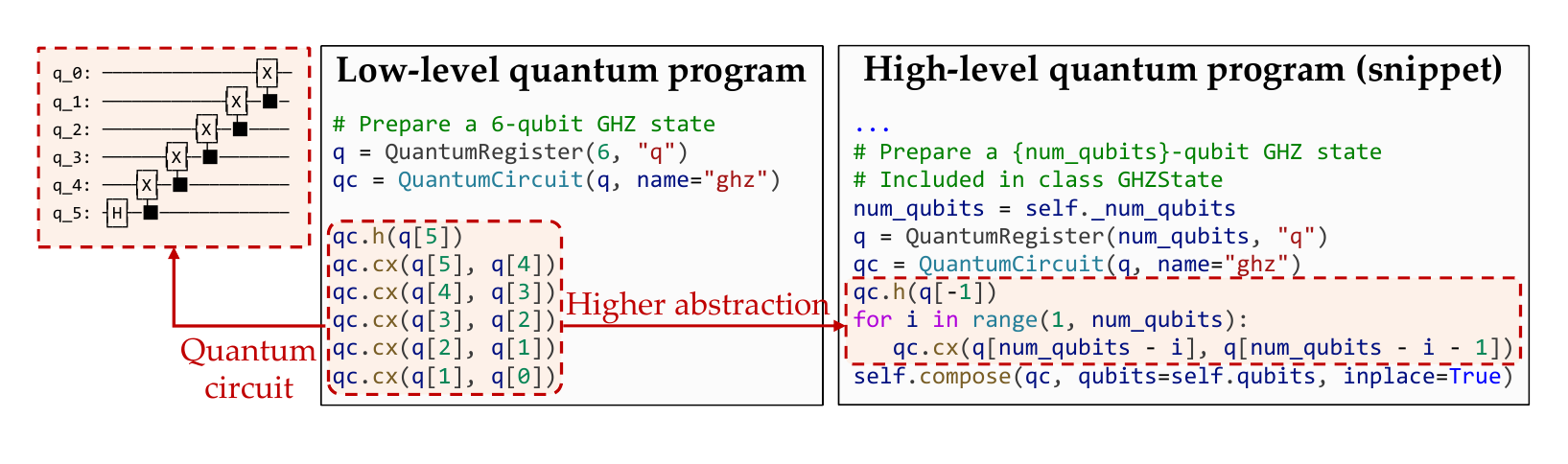}
    \caption{Two different implementations of GHZ state preparation}\label{fig: background}
\end{figure}

Quantum programs are designed to implement quantum algorithms or subroutines. Especially for the mainstream gate-based quantum computation, the computational procedure of a quantum algorithm is realized by a sequence of quantum gates acting on qubits. The pure state of a qubit can be denoted as a state vector $\ket{\psi}$, and a quantum gate can map the qubit state to $U\ket{\psi}$, where $U$ is mathematically a unitary operator. Regarding running quantum programs on physical hardware or their corresponding classical simulation, program outputs are not accessible until they are transformed into probabilistic outcomes through quantum measurement. 

Today, quantum programming languages and frameworks, such as Qiskit, built on the host language Python, provide fruitful APIs for programmers to develop quantum applications. Following the practice of classical software engineering (CSE), these official APIs largely involve modularity and scalability, which requires appropriate code abstractions. 

\revise{To compare code at different abstraction levels, Figure~\ref{fig: background} illustrates two Qiskit examples, both of which realize the  Greenberger-Horne-Zeilinger (GHZ) state preparation.} 
In the left code, each statement after circuit construction corresponds to a concrete quantum gate, but this hard-coded approach limits functionality to preparing a 6-qubit state only. By contrast, the right-hand example from~\BenchmarkName incorporates higher abstraction for quantum gates and is scalable to prepare a general GHZ state, whose qubit number can be flexibly determined by the classical variable \texttt{self.\_num\_qubits}.


\revise{Abstraction hides low-level implementation details behind high-level interfaces, supporting the modularity, reusability, and scalability needed as quantum systems grow in both size and complexity~\cite{zhao2025when}. 
Nevertheless, prior QST studies have paid limited attention to scalable quantum programs with such abstractions, considering only a few exceptions including~\cite{long2024testing, long2025black, li2025preparation, li2016dynamic}. 
Hence, \BenchmarkName addresses this gap by enabling systematic evaluation of testing techniques on scalable quantum programs.}

\subsection{Benchmarks for Quantum Software}

\begin{table}[htbp]
    \footnotesize
    \renewcommand{\arraystretch}{0.9}
    \caption{Comparison among open-source benchmarks for general research on quantum software}\label{tab: benchmark}
    \begin{center}
        \resizebox{\columnwidth}{!}{
            
\begin{tabular}{c|c c C{0.16\columnwidth} C{0.13\columnwidth} c c}
    \toprule[1pt]
    \multicolumn{1}{c|}{\textbf{Name}} & 
    \multicolumn{1}{c}{\textbf{Used?}} & 
    \multicolumn{1}{c}{\textbf{Scalable?}} & 
    \multicolumn{1}{c}{\textbf{Framework}} & 
    \multicolumn{1}{c}{\textbf{Target}} & 
    \multicolumn{1}{c}{\textbf{Component}} & 
    \multicolumn{1}{c}{\textbf{Size}}\\
    \cmidrule(lr){1-1} \cmidrule(lr){2-2} \cmidrule(lr){3-3} \cmidrule(lr){4-4} \cmidrule(lr){5-5} \cmidrule(lr){6-6} \cmidrule(lr){7-7}
    Bugs4Q~\cite{zhao2021bugs4q, zhao2023bugs4q} 
    & \markyes
    & \markno
    & Qiskit
    & Software 
    & Bugs 
    & 36, 42 \\
    \cmidrule(lr){1-1} \cmidrule(lr){2-2} \cmidrule(lr){3-3} \cmidrule(lr){4-4} \cmidrule(lr){5-5} \cmidrule(lr){6-6} \cmidrule(lr){7-7}
    MQT Bench~\cite{quetschlich2023mqt}
    & \markyes
    & \markno*
    & OpenQASM
    & Hardware
    & Algorithms 
    & 28 \\
    \cmidrule(lr){1-1} \cmidrule(lr){2-2} \cmidrule(lr){3-3} \cmidrule(lr){4-4} \cmidrule(lr){5-5} \cmidrule(lr){6-6} \cmidrule(lr){7-7}
    QSimBench~\cite{bisicchia2025qsimbench}     
    & \markno
    & \markno
    & Qiskit  
    & Software  
    & Algorithms 
    & 14 
    \\
    \cmidrule(lr){1-1} \cmidrule(lr){2-2} \cmidrule(lr){3-3} \cmidrule(lr){4-4} \cmidrule(lr){5-5} \cmidrule(lr){6-6} \cmidrule(lr){7-7}
    Lubinski et al.~\cite{lubinski2023application} 
    & \markno
    & \markyes
    & Qiskit, Cirq, Braket, Q\#
    & Hardware
    & Algorithms 
    & 13
    \\
    \cmidrule(lr){1-1} \cmidrule(lr){2-2} \cmidrule(lr){3-3} \cmidrule(lr){4-4} \cmidrule(lr){5-5} \cmidrule(lr){6-6} \cmidrule(lr){7-7}
    SupermarQ~\cite{tomesh2022supermarq} 
    & \markno
    & \markyes
    & Qiskit
    & Hardware, Software
    & Algorithms 
    & 8
    \\
    \cmidrule(lr){1-1} \cmidrule(lr){2-2} \cmidrule(lr){3-3} \cmidrule(lr){4-4} \cmidrule(lr){5-5} \cmidrule(lr){6-6} \cmidrule(lr){7-7}
    VeriQBench~\cite{chen2022veriqbench}
    & \markyes
    & \markno*
    & OpenQASM
    & Software
    & Algorithms 
    & 21
    \\
    \cmidrule(lr){1-1} \cmidrule(lr){2-2} \cmidrule(lr){3-3} \cmidrule(lr){4-4} \cmidrule(lr){5-5} \cmidrule(lr){6-6} \cmidrule(lr){7-7}
    QASMBench~\cite{li2023qasmbench} 
    & \markyes
    & \markno
    & OpenQASM
    & Hardware
    & Algorithms 
    & 35
    \\
    \cmidrule(lr){1-1} \cmidrule(lr){2-2} \cmidrule(lr){3-3} \cmidrule(lr){4-4} \cmidrule(lr){5-5} \cmidrule(lr){6-6} \cmidrule(lr){7-7}
    RevLib~\cite{wille2008revlib}     
    & \markyes
    & \markno
    & SyReC     
    & Hardware  
    & Circuits  
    & 154       
    \\
    \cmidrule(lr){1-1} \cmidrule(lr){2-2} \cmidrule(lr){3-3} \cmidrule(lr){4-4} \cmidrule(lr){5-5} \cmidrule(lr){6-6} \cmidrule(lr){7-7}
    \textbf{\BenchmarkName (ours)}
    & N/A
    & \markyes
    & Qiskit
    & Software
    & Algorithms 
    & \NumberOfPrograms 
    \\
    \bottomrule[1pt]
\end{tabular}

        }
        {\justify\footnotesize
            \revise{``Used?'' indicates whether the benchmark has been applied in existing QST studies. Compared to ``\markno'' in the column of ``Scalable?'' denoting no scalable quantum programs provided, ``\markno*'' means that several, but not most, of the source quantum programs in the benchmark backend are scalable.  ``Size'' corresponds to the component number reported in the literature, combined with multiple levels, like algorithm types, circuit instances and bugs.}
        \par}
    \end{center}
\end{table}

\revise{Recent studies have proposed benchmarks for quantum code, such as QuanBench~\cite{10.1109/ASE63991.2025.00218} and QCircuitBench~\cite{yang2026qcircuitbench}, which are intended to evaluate large language models in code generation.}
\revise{With a focus on underlying use for QST, Table~\ref{tab: benchmark} lists more relevant open-source benchmarks covering QC and SE perspectives.} 
\revise{For the scope beyond assembly-level and fixed-size OpenQASM programs, Qiskit is the most popular framework, which motivates building~\BenchmarkName on top of it.}
In detail, Bugs4Q collects real-world bug-fix pairs in quantum programming, where many bugs do not result from unexpected behavior of quantum programs, but from misuse of quantum software stacks. Meanwhile, RevLib is an early repository of general reversible circuits for computer-aided design, rather than being dedicated to QC. The latest benchmark, QSimBench, is oriented toward QSE research on circuit-level objects and provides precomputed outcomes rather than program code.
Since Table~\ref{tab: benchmark} counts heterogeneous entities, benchmark sizes are not directly comparable, while~\BenchmarkName still has the largest subject count among benchmarks whose primary components are quantum algorithms.

Several benchmarks~\cite{quetschlich2023mqt, lubinski2023application, li2023qasmbench, wille2008revlib} aim to evaluate performance of quantum hardware and its simulation. OpenQASM programs are common as they serve as an assembly-level intermediate representation that bridges high-level quantum algorithms and low-level hardware execution layers. 
Therefore, QC research uses these circuit-like programs as test cases to evaluate hardware performance characteristics such as fidelity under noise and entanglement capability~\cite{li2023qasmbench}.
Nevertheless, developers rarely construct applications at such a low level. \revise{QST focuses on program functionality and bug detection within the QSDLC}. This distinction highlights the difference between testing quantum hardware and testing quantum programs.

\revise{Bugs4Q, MQT Bench, VeriQBench, QASMBench, and RevLib have been used for QST, but scalable quantum programs have been scarcely involved.}
Benchmarks~\cite{lubinski2023application, tomesh2022supermarq} have not been explored for QST but include high-quality scalable programs, making them appropriate sources for~\BenchmarkName to enhance program diversity. In addition, compared to many scalable programs included in existing benchmarks, which only consider input arguments directly mapped to qubit counts,~\BenchmarkName also exposes the input ports of other classical arguments (e.g., the database to be searched in Grover Search), thereby enabling richer program configurations and input spaces for QST experiments.

\section{Methodology of Benchmark Design}\label{sec: methodology}
\subsection{Pipeline of Benchmark Construction}
\begin{figure}[htbp]
    \centering
    \includegraphics[width=0.9\columnwidth]{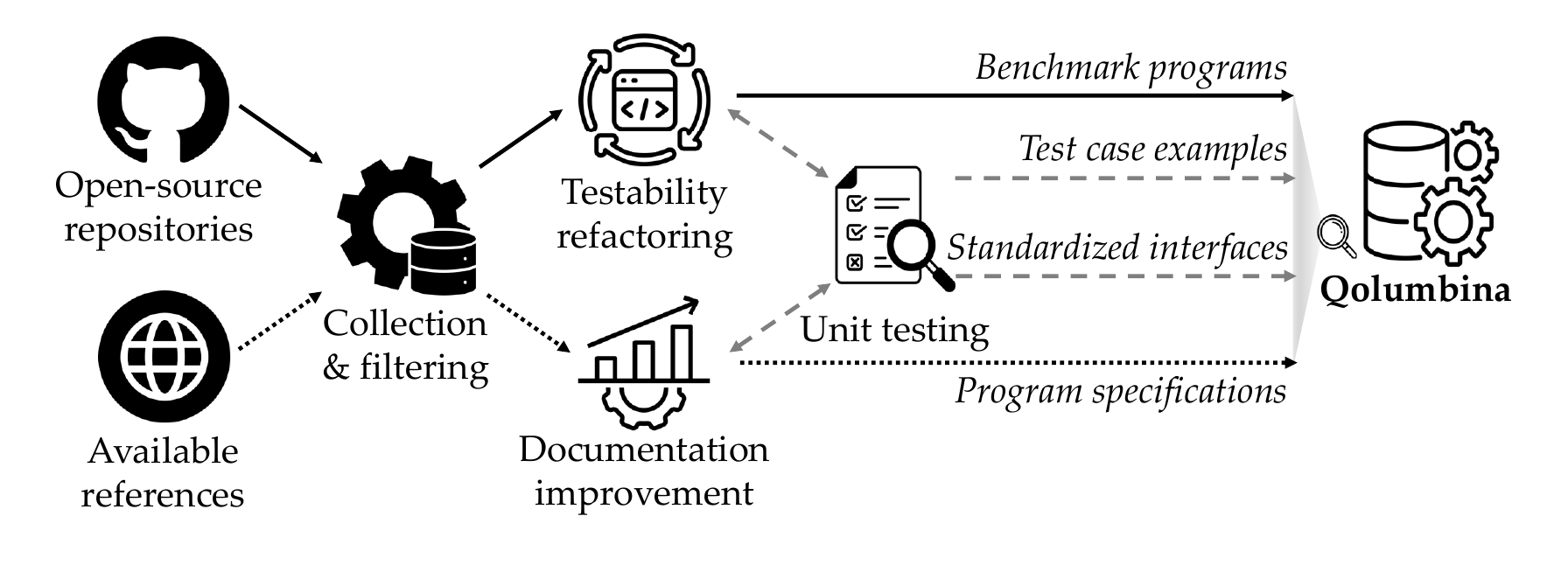}
    \caption{Overview of~\BenchmarkName's construction}\label{fig: framework}
\end{figure}

{\BenchmarkName is a benchmark infrastructure built on Qiskit 2.3.0 to support controlled and reproducible QST experiments. 
Figure~\ref{fig: framework} illustrates a pipeline for its construction and structure, with details provided in the following subsections. 
We first collect open-source quantum software repositories to identify scalable quantum programs. 
The selected programs are then moderately refactored to standardize interfaces and support valid test execution. 
Since many repositories lack clear descriptions of program functionality and usage, we consult accessible literature and tutorials to prepare QST-oriented documentation. 
Finally, we develop unit tests for each benchmark program and iteratively refine both the refactored programs and documentation based on test feedback.}

\subsection{Program Collection and Filtering}
{To promote program diversity while ensuring a reasonable selection scope, we consider three types of accessible sources for benchmark programs {with real-world provenance}:}
\begin{itemize}[leftmargin=*]
    \item \textbf{C1}: Program sources or benchmarks mentioned or considered by QST studies 
    \item \textbf{C2}: Artifacts of QST studies, which explicitly provide original PUTs
    \item \textbf{C3}: Relevant benchmarks cited by the publications that propose the benchmarks marked as \textbf{C1}
\end{itemize}
{Specifically, we use real-world provenance to refer to traceable sources rather than deployment evidence. A program is considered to exhibit real-world provenance if it is claimed in the corresponding research paper, carries copyright information from a recognized quantum software project, or is hosted in an official quantum SDK repository like Qiskit.}
An existing survey~\cite{li2026methodological} summarizes 15 \textbf{C1} sources and 33 \textbf{C2} sources. For~\textbf{C3}, we newly extracted 2 sources, i.e., Lubinski et al.'s benchmark~\cite{lubinski2023application} and SupermarQ~\cite{tomesh2022supermarq}, by revisiting the literature~\cite{quetschlich2023mqt}. 

Then, we filtered out the candidate quantum programs that violated at least one of the following three criteria:
\begin{itemize}[leftmargin=*]
    \item \textbf{F1}: Programs scalable in both logic and structure, {allowing classical arguments to flexibly determine quantum circuits}
    \item \textbf{F2}: Programs accompanied by functional descriptions in their corresponding sources, or whose underlying algorithmic principles are elaborated on in accessible literature
    \item \textbf{F3}: Programs implementing a unitary operation of a quantum algorithm in a noise-free situation
\end{itemize}
\textbf{F1} helps exclude programs that inherently represent fixed-size quantum circuits (e.g., preparing a Bell state with only two qubits), or heavily hard-coded logic (e.g., the left program in Figure~\ref{fig: background}). \textbf{F2} is motivated by the assumption of testability, as test oracles would be infeasible to properly design or systematically derive in the complete absence of knowledge about expected behavior of the program.
\textbf{F3} follows the current focus on testing-related tasks, in which the evolution of the qubit within a quantum circuit is mathematically modeled as a unitary operation $U$~\cite{li2024automatic, oldfield2025faster, miranskyy2025feasibility}.
In fact, a general class of quantum algorithms involving mid-circuit measurements falls outside the scope described above. Nevertheless, the difficulty in accurately depicting the program specification, together with the scarcity of related QST studies~\cite{paltenghi2024survey}, makes it reasonable for~\BenchmarkName to consider these cases as future extensions.

We completed program collection before~\DDL. Using \textbf{F1}--\textbf{F3} as exclusion criteria, and after author discussion and consensus, we retained~\NumberOfPrograms deduplicated programs, treating them as distinct subjects when they exhibited observable structural or functional
differences. To maintain source traceability, we preserved copyright information and repository links for each retained program.
Especially, \textbf{C1} provided 22 programs from sources previously used for testing, where three official Qiskit repositories~\cite{Qiskit-qiskit, qiskit_qiskit-textbook, qiskit-community_qiskit-textbook} and MQT Bench~\cite{quetschlich2023mqt} offered 17 and 5 programs, respectively. 
There were 12 directly included in artifacts of three prior studies~\cite{abreu2022metamorphic, li2025preparation, long2025black} (\textbf{C2}), while Lubinski et al.'s benchmark~\cite{lubinski2023application} contributed the remaining 6 (\textbf{C3}).
To present a compact overview of benchmark programs, we manually organized them into~\NumberOfFamilies descriptive families
according to their documented primary functionality, where all these families and included programs are listed in our artifact~\cite{qolumbina_infrastructure}.
For example, the family \texttt{quantum\_adder} incorporates three types of adders executable in quantum circuits: Draper adder, full adder, and weighted adder.

\subsection{Testability Refactoring}

{Following testability refactoring in
SE practices~\cite{reich2023testability}, we moderately refactored the
collected programs to unify heterogeneous interfaces and support unit testing. {Overall, 95\% of original programs underwent at least one of the following four testability-oriented adaptations with localized code changes intended to improve testability and maintainability while preserving output behavior for valid inputs. 
Modifications for refactoring are documented in code comments of the corresponding programs.
In Section~\ref{sec:
unit_testing}, we will further introduce the use of unit tests to practically mitigate potential functional differences after refactoring.}}
\begin{itemize}[leftmargin=*]
    \item \textbf{Structure reorganization}: Program structures are reorganized to unify input-output ports across our infrastructure.
    \item \textbf{Dependency decoupling}: {External
    dependencies that may affect program functionality or
    evolve over time are replaced with local equivalent
    implementations.}
    \item \textbf{Input validation}: Assertions are inserted to prevent executing invalid test inputs that violate the program specification. 
    \item \textbf{Cross-language translation}: Programs written in other frameworks or languages are translated into Qiskit/Python and validated against their program specifications.
\end{itemize}

{Thirty-seven original programs were structurally
reorganized, including unifying programs into classes
that inherit from Qiskit's~\texttt{QuantumCircuit} and
converting fixed classical variables into configurable
input arguments.}

{Dependency decoupling was performed in
15 programs, following the insights
of~\cite{spadini2019mock}. For example, QFT subroutines
that originally relied on the Qiskit package were
replaced with equivalent implementations provided
in~\BenchmarkName. This improves the long-term usability
and controllability of the benchmark, because the current
QFT implementation is scheduled for deprecation in an
upcoming version of Qiskit~\cite{qiskit_qft-2026} (commit
\texttt{3fe73d9}).}

{Input validation was applied to 13 programs to support sound functional
testing by rejecting invalid test inputs that would
otherwise trigger undefined behavior. Specifically,
constraints were enforced for both physical
interpretability and logical rationality, such as the
non-negativity of the evolution time in Hamiltonian
simulation and the requirement that integers provided to
comparator circuits be representable with the given
number of qubits.}

{Finally, 10 programs originally written
in Q\# were manually translated into Qiskit counterparts conforming to the program specifications given by their source~\cite{long2025black}. 
Since not all implementations have one-to-one counterparts in Qiskit, we constructed equivalent quantum operations when needed; for example, Qiskit's Hadamard and multi-controlled $X$ gates were used to implement Q\#'s multi-controlled $Z$ gate.}

\subsection{Testing-oriented Program Specifications}
The program specifications provided on a webpage~\cite{qolumbina_documentation_link} aim to facilitate controlled QST experiments by elaborating on the input arguments and expected behavior for each benchmark program. 
{These documents are based on source descriptions and
algorithm literature, such as the error-bound analysis of
Quantum Monte Carlo~\cite{woerner2019quantum}, to support test oracle design and semantic checks after refactoring.}
To ensure specification readability, we consider the following five elements:

\textbf{Formula-based specification.} 
According to~\cite{long2024testing, li2026methodological}, we utilized mathematical formulas to formally depict the expected output or structure of each benchmark program. For example, a mapping between state vectors,  
$\ket{y}_n\ket{x}_n \mapsto \ket{(x+y)\text{ mod } 2^n}_n\ket{x}_n$, is introduced for \texttt{draper\_adder} with two $n$-qubit input computational basis states $\ket{x}_n$ and $\ket{y}_n$ $(x,y\in\mathbb{N})$. 
{Meanwhile, for programs with multiple subroutines, such as
\texttt{grover\_search}, we specified subroutine-level semantics,
including the phase oracle, diffusion operator, and iteration-wise
state updates.}
To mitigate confusion in specification interpretation, we underscored the little-endian ordering and qubit counts in the formal representation---details that are sometimes overlooked or insufficiently clarified even in parts of Qiskit's official documentation. 
The little-endian ordering allows us to rigorously specify register indices, such as the above initial state $\ket{y}_n\ket{x}_n\equiv \ket{y_{n-1} \cdots y_0x_{n-1} \cdots x_0}_{2n}$ with the bits $y_{n-1}$ and $x_0$ respectively occupying the most and least significant positions within the length-$2n$ array. Our explicit emphasis on qubit counts is motivated by the two versions \texttt{integer\_comparator\_greedy} and \texttt{integer\_comparator\_old} sourced from Qiskit. Although both provide similar functionality, they employ inconsistent numbers of qubits (i.e., $1$ and $n$, respectively) to encode the comparison result, which may otherwise lead to ambiguity in formal specifications.

\textbf{API documentation.} The API documentation is generated via Sphinx~\cite{sphinx} that automatically parses and formats the docstrings of program modules. The program variables are linked with the symbols denoted in the formula-based specifications. The potentially raised errors are listed to remind testers to avoid invalid inputs. In addition, we noticed that the official Qiskit documentation might omit functional descriptions of some arguments used in our benchmark programs, such as the \texttt{basis} variable in two programs in the \texttt{pauli\_rotations} family. Thus, we carefully reviewed the source code and supplemented the specification for the corresponding variables.

\textbf{Code example.} 
{We provide code examples to facilitate the construction of relatively complex test inputs. For example, \texttt{counting} requires an oracle operation as input, which may be represented either by a state vector or by a quantum circuit. Accordingly, we include two code examples to demonstrate both representations.}

\textbf{Documentation references.} We provide references used to prepare our program specifications, which strengthen traceability and provide supporting materials for testers.

\textbf{Technical debt.} 
{Given the growing attention to technical debt in quantum software~\cite{openja2022technical, ishimoto2024empirical}, we explicitly marked two debts in the documentation to make known limitations and future maintenance
costs visible. 
One is a development debt in~\texttt{multiplier\_hrs}, where some functional arguments remain unexposed due to redesign risks and validation costs. 
The other is a testing debt in~\texttt{monte\_carlo}, where simple and common QST oracles, such as those that compare individual outcomes or full output distributions~\cite{ali2021assessing}, are theoretically unreliable for this approximate algorithm.}

\subsection{Unit Testing}\label{sec: unit_testing}
\begin{figure}[htbp]
    \centering
    \includegraphics[width=\columnwidth]{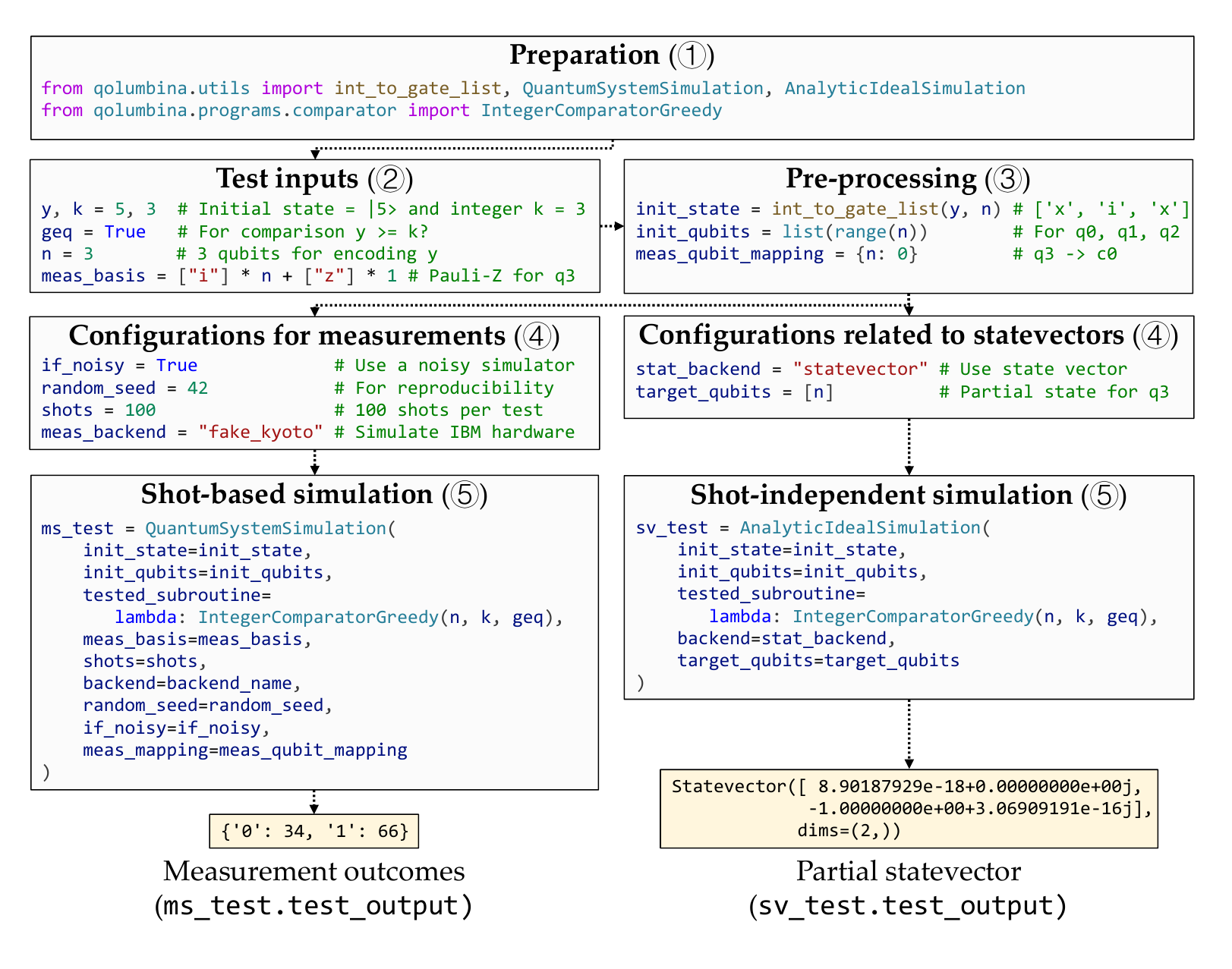}
    \caption{A code example to run the benchmark quantum program~\texttt{integer\_comparator\_greedy} through two backends}\label{fig: code_example}
\end{figure}

\BenchmarkName includes unit tests both as executable examples for follow-up QST research and as specification-based conformance checks. 
These tests exercise the standardized interfaces and check whether the refactored programs conform to the documented input constraints and expected outputs, thereby helping mitigate functional drift induced by refactoring.
For a quasi-blind validation, two authors, independent of refactoring original programs and documenting program specifications, {designed and executed~\NumberOfUnitTests program-level unit tests across the~\NumberOfPrograms benchmark programs}. They provided feedback based on the test results, further improving refactoring quality and program specifications. As a consequence, all provided test cases passed, achieving a statement coverage of~\Coverage. 
This provides practical confidence that the refactored programs conform to the documented specifications on the exercised cases. Objectively, it is unrealistic to establish semantic equivalence to all original behaviors because the refactoring itself introduces reasonable changes to refine program testability.

{In detail, all tests were executed through standardized interfaces provided by~\BenchmarkName, as illustrated in Figure~\ref{fig: code_example}. 
These interfaces expose a subroutine under test, classical arguments, quantum inputs and measurements, and execution configurations, like shots, backends, random seeds, and output qubits. 
This higher-abstraction interface design simplifies test construction by hiding low-level Qiskit API details.}

Both shot-based and shot-independent backends support test execution. 
The shot-based one, following the measurement-based execution commonly assumed in QST studies, supports one ideal simulator and 60 fake backends. The ideal simulator mainly targets theoretical analysis of computational procedure, while fake backends can better approximate characteristics of real quantum hardware.
The shot-independent one, rather than simulating execution on physical quantum hardware, produces statevectors or unitary operators to enable direct checks of quantum states or transformations; this has recently gained attention in research~\cite{miranskyy2025feasibility} and is also used in Qiskit testing practice~\cite{Qiskit-qiskit}.


\section{Empirical Study}\label{sec: empirical}
\subsection{Overview of Research Questions}
\revise{To characterize testing-related properties of scalable
quantum programs and outline implications for
controlled experiments, we propose the following
research questions (RQs):}

\begin{enumerate}[label=\textbf{RQ\arabic*}, leftmargin=*]
    \item What are benchmark programs' functional properties?
 
    \textit{RQ1.1} To what extent do the programs with real-world provenance go beyond purely pedagogical and artificial instances?
    
    \textit{RQ1.2} What are the output characteristics of programs?
 
    \item What are the scales of benchmark programs?

    \textit{RQ2.1} What is the development complexity of the programs?  
    
    \textit{RQ2.2} What is the execution complexity of the programs?  
    
    \item What is the performance of adopting benchmark programs for controlled QST experiments?
 
    \textit{RQ3.1} What is the time cost of program execution?
    
    \textit{RQ3.2} What is the performance of employing benchmark programs for fault detection? 
\end{enumerate}

\revise{Motivated by QST's goal of checking the functional correctness of PUTs, RQ1 characterizes benchmark functionality through real-world provenance and output characteristics. 
Concerning that prior QST studies often used tutorial-style instances~\cite{li2026methodological}, mostly designed around artificial problems to demonstrate quantum advantages, RQ1.1 examines whether~\BenchmarkName includes programs beyond pedagogical or artificial instances, thereby supporting empirical research on more practically relevant subjects. 
RQ1.2 investigates program outputs, since interpreting test outcomes requires knowing what expected output should be compared against observed behavior, which directly affects test oracle design.}

Test scalability is one crucial requirement that expects the test approaches to still work for large-scale and sophisticated programs. 
\revise{To this end, RQ2 evaluates whether~\BenchmarkName provides suitable PUTs for studying test scalability. 
RQ2.1 assesses development complexity, as developers are generally more likely to introduce defects in programs with complex structures. RQ2.2 complements this view from the execution perspective by quantifying quantum circuit scale, a commonly considered factor in existing QST studies.}

\revise{RQ3 conducts two controlled experiments using \BenchmarkName on shot-based simulators to evaluate whether it supports methodology of existing QST empirical studies. 
In terms of test cost and effectiveness, we reuse MSTC~\cite{li2025preparation} and DOSS~\cite{li2016dynamic} for feasibility analysis of \BenchmarkName because they are recent QST approaches with both available and functional artifacts, along with prior evaluations on scalable Qiskit programs.
Across different backends, RQ3.1 follows MSTC by comparing execution time between two test-suite designs, while RQ3.2 uses DOSS as a comparably reliable test oracle to study reported failures on buggy benchmark variants.}
 
\subsection{RQ1: Functional Analysis}
\subsubsection{Data Annotation}
\begin{table}[htbp]
    \caption{Introduction of data annotation}
    \begin{center}
        \label{tab: annotation}
        \footnotesize
        \resizebox{.98\columnwidth}{!}{
            
\begin{tabular}{c|p{0.76\columnwidth}|p{0.23\columnwidth}}
    \toprule[1pt]
    \multicolumn{1}{c|}{\textbf{Task}} & \multicolumn{1}{c|}{\textbf{Description}} & \multicolumn{1}{c}{\textbf{Return}} \\
    \cmidrule(lr){1-1} \cmidrule(lr){2-2} \cmidrule(lr){3-3}  
    \textbf{T1} &
    \revise{Determines whether the quantum algorithm of a program is primarily instructional. For example, an algorithm that often occurs in official tutorials or learning modules, and mainly illustrates theoretical principles and algorithmic advantages rather than practical usage is marked as ``Yes''.}
    &
    A Boolean tag (``Yes'' or ``No'') \\
    \cmidrule(lr){1-1} \cmidrule(lr){2-2} \cmidrule(lr){3-3}  
    \textbf{T2} &
    \revise{Identifies whether a quantum program is reusable as a subroutine of larger quantum algorithms or applications. Such a program does not implement a fully standalone function or complete application; instead, it is designed modularly, making it easy to integrate into a larger circuit.} 
    &A Boolean tag (``Yes'' or ``No'') \\
    \cmidrule(lr){1-1} \cmidrule(lr){2-2} \cmidrule(lr){3-3}  
    \textbf{T3} &
    Investigates what application domains the quantum program is designed for. &
    At least one candidate category\\
    \cmidrule(lr){1-1} \cmidrule(lr){2-2} \cmidrule(lr){3-3}  
    \textbf{T4} &
    Investigates what type of solution-related output is expected by solely executing a quantum program upon a simulated or physical quantum system. &
    At least one candidate category \\
    \bottomrule[1pt]
\end{tabular}
        

        }
    \end{center}
\end{table}

Three authors with strong expertise in QSE performed the category annotation of each benchmark program in terms of their functionalities. 
Table~\ref{tab: annotation} lists the four involved tasks, where \textbf{T1}, \textbf{T2}, and \textbf{T3} are proposed to answer RQ1.1, while \textbf{T4} is considered especially for RQ1.2.

\revise{For~\textbf{T3} and~\textbf{T4} involving predefined categories,
since no existing taxonomy was found to match such analysis for QST, we carefully reviewed program specifications and source codes to determine the candidate categories along with their definitions.
Then, three authors independently annotated all the~\NumberOfPrograms programs, following a clear guideline provided in our artifact~\cite{qolumbina_empirical}. 
Fleiss' $\kappa$~\cite{fleiss1971measuring} is used to measure the agreement among authors beyond two.
The agreement scores for \textbf{T1}--\textbf{T4} are respectively 0.77, 0.70, 0.73, and 0.81. For the multi-label tasks \textbf{T3} and \textbf{T4}, we compute Fleiss' $\kappa$ on each binary label and report the support-weighted macro average. According to~\cite{landis1977measurement}, all scores indicate substantial agreement. Furthermore, the authors organized an online discussion to resolve disagreements and then decide on the annotated metadata.}

\subsubsection{RQ1.1: Beyond Pedagogical and Artificial Instances}
\begin{figure}[htbp]
    \centering
    \subfloat[\textbf{T1}: Tutorial identification\label{fig: tutorial}]{%
        \includegraphics[width=0.48\columnwidth]{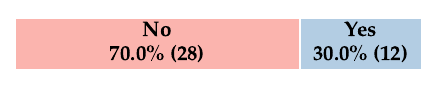}
    }%
    \hfill
    \subfloat[\textbf{T2}: Reusability identification\label{fig: subroutine}]{%
        \includegraphics[width=0.48\columnwidth]{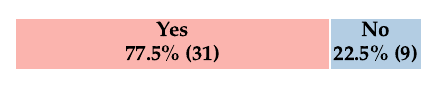}
    }

    \subfloat[Categories for program functionalities (\textbf{T3})\label{fig: function}]{%
        \includegraphics[width=0.88\columnwidth]{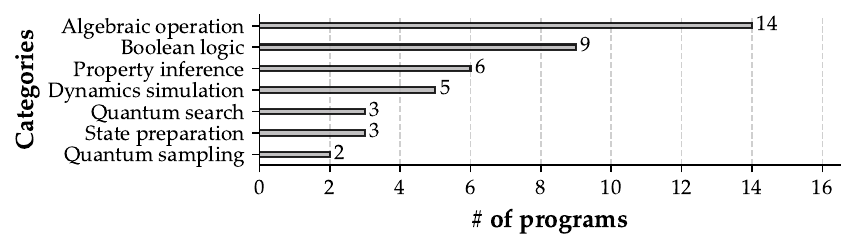}
    }
    \caption{Analysis of programs with real-world provenance}
    \label{fig: applicability}
\end{figure}

As shown in Figure~\ref{fig: tutorial}, the majority (i.e., 70\%) of the benchmarked algorithms are not primarily pedagogical.
Hence, moving beyond purely pedagogical and artificial benchmarks allows QST approaches to be evaluated on programs with clearer program-level functionality and real-world provenance.

\revise{For program instances, Figure~\ref{fig: subroutine} exhibits that $77.5\%$ of benchmark programs could be reused as a subroutine. 
This indicates that many programs capture modular components that can appear in larger quantum workflows, rather than only standalone demonstrations.
Several reusable instances also implement non-tutorial algorithms.
For example, oracle programs in the \texttt{quantum\_adder} and \texttt{quantum\_multiplier} families implement reversible arithmetic primitives, which can be integrated into larger quantum algorithms, including applications of quantum finance such as credit risk assessment~\cite{egger2020quantum}.}


\revise{Figure~\ref{fig: function} shows that the benchmark incorporates a diverse spectrum of seven functional categories.
The largest category contains 14 programs for algebraic or arithmetic operations on quantum registers, followed by 9 programs for classical-predicate evaluation. 
The remaining programs cover inference of hidden properties and structures, simulation of quantum dynamics, quantum-based search algorithms, state preparation subroutines, and sampling-related estimation, including representative cases such as Simon's algorithm, Hamiltonian Simulation, Grover search, graph-state preparation, amplitude estimation, and quantum Monte Carlo. 
This distribution helps reduce the risk of evaluating QST techniques on a narrow set of algorithmic behaviors.}

\AnswerToRQ{1.1}{\revise{Most benchmark programs are non-pedagogical ($70.0\%$) and reusable as subroutines ($77.5\%$). 
Their functionalities span seven heterogeneous categories, with algebraic operations being the largest. 
These results suggest that \BenchmarkName includes a substantial proportion of programs with development-relevant characteristics while maintaining functional diversity.}}

\subsubsection{RQ1.2: Output Characteristics}

Motivated by prior studies~\cite{oldfield2025faster, long2025black, li2016dynamic} that test quantum programs with specific output properties, we summarize six output types to characterize the intended outputs of quantum programs. 
We assign each category based on the specified measurement on output qubits associated with program functionality, while we assume the default Pauli-$Z$ basis when no measurement is predefined.
Besides, one program may belong to multiple categories, such as QFT storing outputs in both amplitudes and phases, making it classifiable as ``exact distribution output'' and ``phase-encoded output''.

\begin{table*}[!t]
    \caption{Categories for output characteristics (\textbf{T4})}
    \begin{center}
        \label{tab: output_category_table}
        \footnotesize
        \renewcommand{\arraystretch}{0.75}
        \resizebox{.98\textwidth}{!}{
            
\begin{tabular}{p{0.08\textwidth}|c|p{0.3\textwidth}|p{0.6\textwidth}}
    \toprule[1pt]
    \multicolumn{1}{c|}{\textbf{Category}} & \multicolumn{1}{c|}{\textbf{\#}} & \multicolumn{1}{c|}{\textbf{Definition}} & \multicolumn{1}{c}{\textbf{Program example}} \\
    \cmidrule(lr){1-1} \cmidrule(lr){2-2} \cmidrule(lr){3-3} \cmidrule(lr){4-4}
    Deterministic output & 15 & 
        The program is expected to yield a deterministic 
        outcome with probability 1 by measuring an eigenstate.
         & 
        \texttt{{draper\_adder}}: Given two $n$-qubit computational
        basis states $\ket{{x}}_n$ and $\ket{{y}}_n$, the program 
        performs the mapping $\ket{{y}}_n \ket{{x}}_n \mapsto 
        \ket{{(x+y) \mod 2^n}}_n \ket{{x}}_n$. This indicates that 
        a correct program yields a measurement outcome with probability 1.
         \\
    \midrule
    Exact distribution output & 11 & 
        The intended output is characterized by
        an exact overall probability distribution rather than a
        few dominant outcomes, meaning the need to observe the amplitudes of a 
        non-eigenstate.
         & 
        \texttt{{w\_state}}: The program prepares an $n$-qubit W state as
        $\ket{{W}}_n = \frac{{1}}{{\sqrt{{n}}}}(\ket{{100\cdots0}} + 
        \ket{{010\cdots0}} + \cdots + \ket{{000\cdots1}})$. With the default
        Pauli-Z measurement, the output distribution upon measuring the state is
        uniform over the $n$ basis states with a single ``1'' and $(n-1)$ ``0''s, 
        and zero for all other basis states.
         \\
    \midrule
    Phase-encoded output & 11 & 
        Output data are encoded in the global or relative
        phase of a quantum state and may not be fully
        observable under common measurement bases.
         & 
        \texttt{{is\_two\_power\_phase}}: Given an $n$-qubit computational basis state
        $\ket{{y}}_n$, this phase-version program computes a Boolean function
        $\ket{{y}}_n \mapsto e^{{i\pi\mathrm{{IS2POWER}}(y)}} \ket{{y}}_n$,
        where the predicate function $\mathrm{{IS2POWER}}(y)$ returns 1 
        if and only if $y$ is a power of 2, and 0 otherwise.
         \\
    \midrule
    Approximate-value dominant output & 4 & 
        Each intended output is concentrated around a dominant region 
        approximating the ideal value, with deviation bounded by a known 
        error tolerance determined by algorithmic parameters.
         & 
        \texttt{{phase\_estimation}}: Given a unitary operator $U$ and
        an eigenstate $\ket{{\psi}}$ of $U$, the program aims to
        estimate the phase $\phi$ using $t$ counting qubits associated with the
        estimation precision. If $2^t \phi$ is an integer, the ideal action 
        of the phase estimation circuit performs $\ket{{\psi}}_n \ket{{0}}_t 
        \mapsto \ket{{\psi}}_n \ket{{2^t \phi}}_t$. Owing to $2^t \phi$ generally being a non-integer, the measurement values $x\in\mathbb{N}$ near 
        $2^t \phi$ will demonstrate dominant peaks in the output distribution.
         \\
    \midrule
    Exact-value dominant output & 2 & 
        The intended output consists of one or more exact measurement values 
        whose probabilities dominate unexpected outcomes.
         & 
        \texttt{{grover\_search}}: Given the set of target items $T$ in a search
        space of size $2^n$, the program approximately returns the output state 
        $\ket{{\psi}}_n = \frac{{1}}{{\sqrt{{|T|}}}} \sum_{{x \in T}} \ket{{x}}_n$ with 
        appropriate iterations. After measurement, each dominant peak corresponding
        to measurement value $x\in\mathbb{N}$ exactly indicates one index of the 
        target items in $T$.
         \\
    \midrule
    Approximate distribution output & 1 & 
        The intended output is characterized by
        an overall probability distribution resulting from an 
        approximate non-eigenstate.
         & 
        \texttt{{hhl}}: Given a linear system
        $A\vec{{x}} = \vec{{b}}$, the program prepares a quantum state
        approximating the normalized solution $\ket{{x}} \propto
        A^{{-1}}\ket{{b}}$, conditioned on measuring the ancilla qubit
        as $1$.
         \\
    \bottomrule[1pt]
\end{tabular}

        }
    \end{center}
\end{table*}

\revise{In Table~\ref{tab: output_category_table}, \BenchmarkName covers six categories, including dominant-output categories that have mentioned in only a few studies~\cite{long2024testing, mendiluze2025quantum} but have earned increasing attention, as well as approximation-related categories that have not been systematically investigated in QST so far. 
Deterministic outputs form the largest category, where testing such quantum programs is close to testing non-randomized classical programs, suggesting opportunities to adapt established CSE testing techniques to QST. 
For the widely studied ``exact distribution output'', \BenchmarkName includes 11 programs and thus supports evaluation on a broader set of scalable programs.
Eleven programs have phase-encoded outputs, posing a challenge for QST because validating their intended behavior may require phase information that is not directly reflected in output probabilities. 
This limitation can make existing test oracles based on hypothesis testing or statistical distance unreliable for phase-encoded programs like QFT~\cite{li2016dynamic}. 
Thus, \BenchmarkName provides useful subjects for evaluating and improving QST techniques for phase-related output behavior.}

\revise{Although fewer programs belong to the remaining categories, they remain valuable for QST research and may expose less-studied testing challenges. 
Both approximate-value and exact-value outputs require test oracle design to focus on functionality-relevant partial outcomes rather than the full output distribution.
Approximate values and distributions further require checking whether errors exceed tolerable bounds.}

\AnswerToRQ{1.2}{\revise{\BenchmarkName covers all six output-related categories, enabling QST techniques to be evaluated across heterogeneous output semantics and oracle requirements. 
Deterministic exact outputs form the largest group, accounting for 37.5\% of the benchmark programs (15 of~\NumberOfPrograms).
Programs with dominant-output or approximation-related outputs, despite being less studied previously, are also included to provide valuable subjects for future research.}}

\subsection{RQ2: Scale Analysis}
\subsubsection{Complexity Metrics}
Referring to Zhao~\cite{zhao2021some}, both complexity metrics general in SE and specific to QSE are considered to quantify the programs scale. 

\revise{RQ2.1 estimates development effort using LOC and cyclomatic complexity, which respectively capture code size and control-flow complexity in a complementary manner. 
LOC counts only effective code lines, excluding blank lines and comments. 
We further exclude dunder methods and factory functions to focus on code relevant to functionality.}

\revise{In the view of quantum circuits corresponding to PUTs, RQ2.2 estimates execution complexity using all the circuit width (a.k.a. number of qubits), size (a.k.a. number of quantum gates), and depth, following prior QST studies~\cite{li2025preparation, li2026methodological, oldfield2025faster}. 
These metrics are computed on the circuits instantiated by classical inputs and transpiled for an ideal backend with full qubit connectivity and a sufficiently expressive basis-gate set, enabling fair comparison across abstracted and synthesized unitary operations.}

\subsubsection{RQ2.1: Development Complexity}
\begin{figure}[htbp]
    \centering
    \subfloat{%
        \includegraphics[width=0.38\columnwidth]{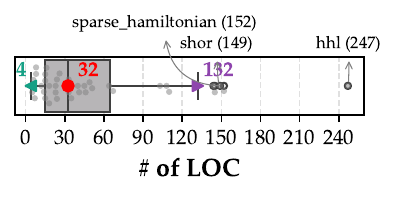}
    }%
    \subfloat{%
        \includegraphics[width=0.42\columnwidth]{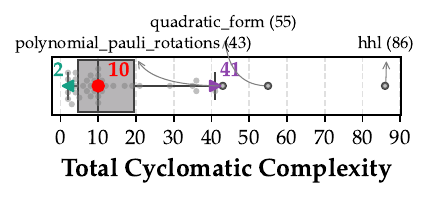}
    }

    \subfloat{%
        \includegraphics[width=0.6\columnwidth]{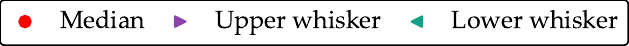}
    }
    \caption{Complexity measures for development effort}
    \label{fig: dev_complexity}
\end{figure}


\revise{Two boxplots in Figure~\ref{fig: dev_complexity} illustrate that \BenchmarkName spans a broad range of development complexity, with LOC ranging from 4 to 247 and cyclomatic complexity from 2 to 86.
Although the median LOC is 32, such compact source codes do not necessarily indicate small testing subjects. 
Unlike low-level quantum programs prevailing in existing benchmarks and studies, whose LOC often reflects gate-level enumeration in one concrete circuit, programs in~\BenchmarkName leverage higher abstractions and classical arguments to generate families of variable circuits. We will analyze instantiated circuit scale separately in RQ2.2.}
  
\revise{Furthermore, five outliers representing large scales are identified, among which the Harrow--Hassidim--Lloyd algorithm (\texttt{hhl}) is the most complicated, possibly attributed to the combination of multiple quantum subroutines.
In general, such outliers are valuable for QST research, as their larger source size and higher control-flow complexity create more implementation paths and subroutine interactions, providing PUTs for examining whether testing techniques scale to structurally complex quantum programs.}

\AnswerToRQ{2.1}{\revise{\BenchmarkName spans a broad range of source-level code sizes and structural complexities, as measured by LOC and cyclomatic complexity. This diversity supports evaluation across programs with different development complexity. Five outliers exposed by at least one metric, especially \texttt{hhl}, provide useful PUTs for studying QST on complex quantum programs in the code level.}}

\subsubsection{RQ2.2: Execution Complexity}
\begin{figure}[htbp]
    \centering
    \subfloat{%
        \includegraphics[width=0.25\columnwidth]{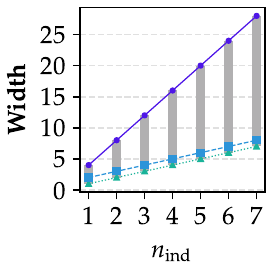}
    }%
    \subfloat{%
        \includegraphics[width=0.25\columnwidth]{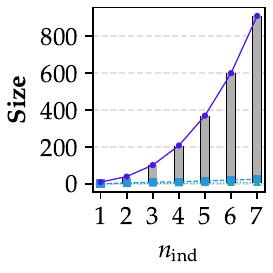}
    }%
    \subfloat{%
        \includegraphics[width=0.25\columnwidth]{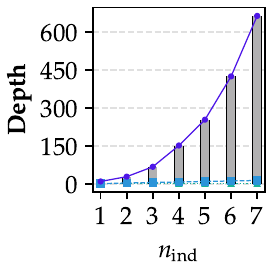}
    }

    \subfloat{%
        \includegraphics[width=0.51\columnwidth]{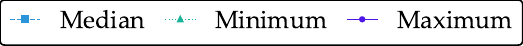}
    }
    
    \caption{Complexity measures for circuit execution}\label{fig: exe_complexity}
\end{figure}
\revise{In RQ2.2, we instantiate quantum circuits by providing each program interface with concrete classical arguments, many of which are reused from the unit tests described in Section~\ref{sec: unit_testing}. In Figure~\ref{fig: exe_complexity}, we examine how a classical argument (denoted as $n_{\mathrm{ind}}$), which directly and independently determines the qubit count of certain quantum registers, affects instantiated circuit scale. Only~\NumberOfProgramsInRQTwoTwo benchmark programs are filtered for investigation, because such an argument $n_{\mathrm{ind}}$ is absent in the others. 
Across the three charts, circuit scale increases with $n_{\mathrm{ind}}$, indicating that \BenchmarkName can support scalability evaluation by varying test inputs. 
This differs from many low-level quantum programs used before, whose fixed circuit structure makes circuit scale largely irrelevant to test inputs.
More specifically, circuit width grows roughly linearly with $n_{\mathrm{ind}}$ across the selected programs and is often larger than $n_{\mathrm{ind}}$ due to automatic inclusion of ancilla qubits or quantum registers. 
By contrast, the maximum circuit size and depth grow much faster with $n_{\mathrm{ind}}$, suggesting that testing larger instantiated circuits on classical simulators can become costly even for small increases in $n_{\mathrm{ind}}$.}


\begin{table}[htbp]
    \caption{Case study of \texttt{weighted\_adder} with LOC and cyclomatic complexity of 144 and 36, respectively}
    \begin{center}
        \label{tab: exe_case_study}
        \footnotesize
        \renewcommand{\arraystretch}{0.75}
        \resizebox{.5\columnwidth}{!}{
            
\begin{tabular}{c|ccc}
    \toprule[1pt]
    \multicolumn{1}{c|}{\textbf{Test inputs $(n_{{\mathrm{{ind}}}}, \mathbf{{w}})$}} & \multicolumn{1}{c}{\textbf{Width}} & \multicolumn{1}{c}{\textbf{Size}} & \multicolumn{1}{c}{\textbf{Depth}} \\
    \cmidrule(lr){1-1} \cmidrule(lr){2-2} \cmidrule(lr){3-3} \cmidrule(lr){4-4}
    $\left(5, [0, 0, 0, 0, 0]^{\top}\right)$ & 6 & 0 & 0 \\
    $\left(5, [1, 3, 5, 7, 9]^{\top}\right)$ & 15 & 680 & 570 \\
    $\left(5, [11, 13, 15, 17, 19]^{\top}\right)$ & 19 & 1,125 & 950 \\
    $\left(6, [0, 0, 0, 0, 0, 0]^{\top}\right)$ & 7 & 0 & 0 \\
    $\left(6, [2, 4, 6, 8, 10, 12]^{\top}\right)$ & 18 & 1,053 & 912 \\
    $\left(6, [14, 16, 18, 20, 22, 24]^{\top}\right)$ & 20 & 1,327 & 1,140 \\
    \bottomrule[1pt]
\end{tabular}

        }
    \end{center}
\end{table}


\revise{To explore other types of classical arguments that could impact the circuit scale, we conduct a case study on \texttt{weighted\_adder}, whose formula-based specification is $\ket{0}_{s}\ket{y}_{n_{\mathrm{ind}}} \mapsto \ket{\mathbf{w}^{\top}\mathbf{y}}_{s}\ket{y}_{n_{\mathrm{ind}}}$. 
Here, $\mathbf{w}=[\lambda_0,\cdots, \lambda_{n_{\mathrm{ind}}-1}]^{\top}$ is an integer vector of $n$ integer weights, the Boolean vector $\mathbf{y}$ encodes integer $y$, and the integer $s = 1+\lfloor \log_2(\sum_{j=0}^{n_{\mathrm{ind}}-1}\lambda_j) \rfloor$ denotes the number of sum qubits. 
Thus, in theory, circuit complexity depends not only on $n_{\mathrm{ind}}$ but also on the weight vector $\mathbf{w}$.
Table~\ref{tab: exe_case_study} exhibits this effect under six test inputs, and we can see that the circuit width, size, and depth are largely governed by specific $\mathbf{w}$. 
For instance, with $n=5$, circuit size dramatically varies from 0 to more than 1,000 across different weight vectors. 
This case highlights that circuit complexity of scalable quantum programs depends not only on the argument with explicit relevance to the qubit count (like $n_{\mathrm{ind}}$), but also on other functionality-related arguments (like $\mathbf{w}$) that would substantially reshape the instantiated circuit, both of which should be carefully considered in test design.}


\AnswerToRQ{2.2}{\BenchmarkName supports testing with large-scale circuits.
Unlike low-level programs with fixed circuits, scalable benchmark programs can produce circuits whose width, size, and depth vary substantially with inputs, including arguments that specify quantum registers or dominate program functionalities.}

\subsection{RQ3: Experimental Analysis}
\subsubsection{Experiment Design}
PUT selection follows the scopes of MSTC~\cite{li2025preparation} and DOSS~\cite{li2016dynamic}. 
RQ3.1 evaluates~\NumberOfProgramsInRQThreeOne programs that functionally support quantum state initialization with computational basis states, enabling comparison between test cases composed of mixed states and pure states with configurations suggested by MSTC.
RQ3.2 runs~\NumberOfProgramsInRQThreeTwo programs categorized with deterministic outputs or exact distribution outputs (referring to Table~\ref{tab: output_category_table}), which matches DOSS's investigation on individual outcomes or overall distribution.
For feasibility analysis on fault detection, seven mutants are generated per program by manually applying mutation operators that correspondingly cover all existing QST mutation actions on quantum gates: adding, deleting, duplicating, modifying, moving, replacing, and switching~\cite{li2026methodological}. 
The code line for mutation is randomly chosen, and human review is performed to validate that mutation operators are not equivalent and indeed alter original functionalities. In RQ3.2, reported failures are caused by incorrect functions instead of program crash.

\revise{Test cases mostly reuse examples from unit tests, along with moderate extensions for enhanced diversity. 
Following MSTC, RQ3.1 compares pure-state and mixed-state test suites with input-domain equivalence, associated with the number of qubits encoding test inputs (denoted as $n_{\mathrm{test}}$). 
Mixed-state test cases require extra qubits for state preparation, but reduce test-suite size while preserving input coverage by grouping multiple pure states into one mixed state.
Overall, RQ3.1 generates \NumberOfMixedTestsInRQThreeOne mixed-state, while RQ3.2 includes programs without the need of state initialization and executes \NumberOfCircuitsInRQThreeTwo instantiated circuits under test. 
Also, test case design considers running large-scale circuits in the context of scalability, as the circuits in RQ3.1 and RQ3.2 respectively have widths up to \MaximumWidthOfRQThreeOne and \MaximumWidthOfRQThreeTwo transpiled on the ideal simulator.} 

Shot-based backends are used throughout RQ3 following common QST practice~\cite{li2026methodological}. Each test execution uses Qiskit's default 1,024 shots and 20 independent repeats with different random seeds to balance statistical stability and execution cost. We evaluate two fake backends supported by \BenchmarkName, Algiers and Brooklyn, under both noiseless and noisy simulation settings, which allows us to study hardware-related impact on QST under affordable simulation costs. Both backends emulate IBM hardware with different native gates and coupling maps, supporting up to 27 and 64 qubits, respectively. Generally, RQ3 experiments are executed on Ubuntu x86\_64 with an Intel(R) Core(TM) i9-10940X CPU @ 3.30GHz.

\subsubsection{RQ3.1: Execution Time}
\begin{figure}[htbp]
    \centering
    \makebox[\columnwidth][c]{%
        \subfloat{%
            \raisebox{-.5\height}{\includegraphics[width=0.48\columnwidth]{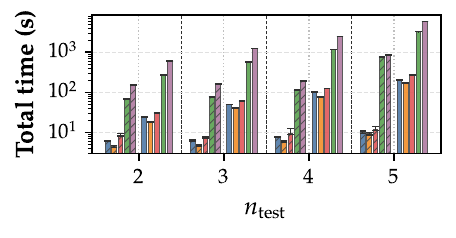}}%
        }%
        \subfloat{%
            \raisebox{-.5\height}{\includegraphics[width=0.2\columnwidth]{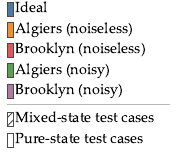}}%
        }%
    }
    \caption{Total time of executing all the test cases of all the involved quantum programs, displayed in a logarithmic axis}\label{fig: time}
\end{figure}
Figure~\ref{fig: time} reports the total execution time of all test cases for each $n_{\mathrm{test}}$. 
In the measured range, execution time increases rapidly and appears approximately linear on the logarithmic scale.
On the three noiseless backends, we reproduce MSTC's finding that pure-state test suites take longer to execute than their mixed-state counterparts.
Beyond MSTC's setting of ideal simulation, we further observe the same trend on noisy simulation, where mixed-state test cases retain their advantage in better efficiency.
Moreover, when running the same test case, noisy backends consistently require longer simulation time than noiseless backends.


\revise{Regardless of simulated noise, we observe clear differences in execution time among three backends.
Across all cases, execution on the ideal backend is not always the fastest, and Brooklyn consistently incurs the longest time, indicating that test overhead can vary substantially across heterogeneous backends.
One plausible explanation is that backend-specific properties, such as qubit connectivity and native gates, affect the transpiled circuit structure and thus simulation efficiency.}


\AnswerToRQ{3.1}{\BenchmarkName reproduces MSTC's finding that mixed-state test cases are more efficient than pure-state counterparts, and extends this observation to noisy simulation.
The three backends also show markedly different execution time, suggesting that backend choice can materially affect QST cost.}

\subsubsection{RQ3.2: Fault Detection}
 
\afterpage{%
    \begin{landscape}
    \begin{figure}[p]
        \centering
        \subfloat{%
            \includegraphics[width=0.55\linewidth]{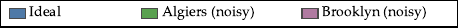}%
        }\par
        \subfloat{%
            \includegraphics[width=0.99\linewidth]{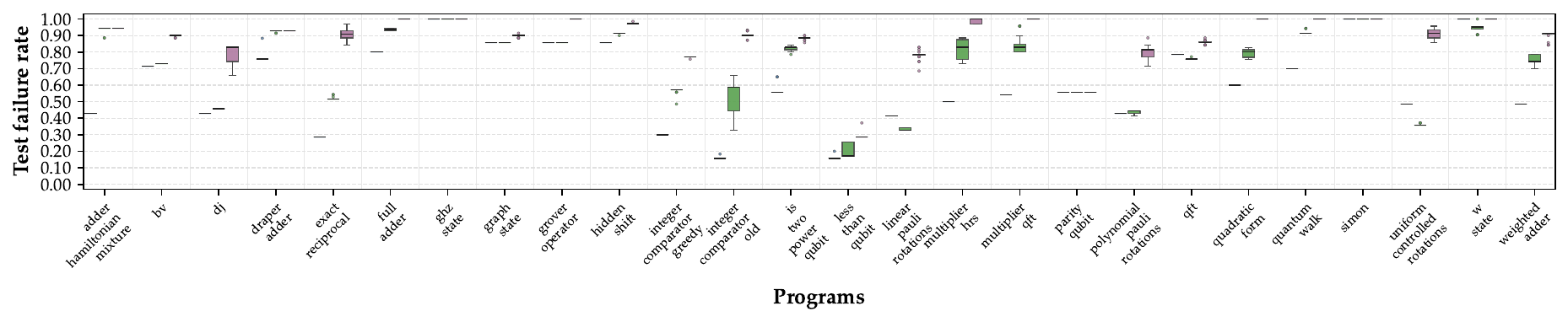}%
        }
        \caption{Boxplot representing test failure rates for \NumberOfProgramsInRQThreeTwo benchmark programs}\label{fig: faults}
    \end{figure}
    \end{landscape}%
    }

\begin{table}[htbp]
    \caption{Statistical analysis of test failure rates across backends}
    \begin{center}
        \label{tab: statistical_faults}
        \renewcommand{\arraystretch}{0.75}
        \resizebox{0.85\columnwidth}{!}{
            
\begin{tabular}{c|cc|cccc}
    \toprule[1pt]
    \textbf{Comparison} & \textbf{$p$-value} & \textbf{$\hat{A}_{12}$} & \textbf{\#N} & \textbf{\#S} & \textbf{\#M} & \textbf{\#L} \\
    \cmidrule(lr){1-1} \cmidrule(lr){2-2} \cmidrule(lr){3-3} \cmidrule(lr){4-4} \cmidrule(lr){5-5} \cmidrule(lr){6-6} \cmidrule(lr){7-7}
    Ideal vs. Algiers & [4.68e-10, 1.00e+00] & [0.00, 1.00] & 5 & 0 & 0 & 21 \\
    Algiers vs. Brooklyn & [4.68e-10, 1.00e+00] & [0.00, 0.50] & 5 & 0 & 0 & 21 \\
    Ideal vs. Brooklyn & [4.68e-10, 1.00e+00] & [0.00, 0.50] & 4 & 0 & 0 & 22 \\
    \bottomrule[1pt]
\end{tabular}

        }
        {\justify\footnotesize
            The $p$-value and $\hat{A}_{12}$ columns report ranges over the discussed programs. The notations \#N, \#S, \#M, and \#L denote the numbers of cases where $\hat{A}_{12}$ represents negligible, small, medium, and large effect sizes, respectively.
        \par}
    \end{center}
\end{table}

  
We use the test failure rate to measure fault-detection effectiveness, defined as the proportion of test executions reported as failed for each PUT.
Figure~\ref{fig: faults} demonstrates the distribution of test failure rate for \NumberOfProgramsInRQThreeTwo programs when DOSS is used as the test oracle on three backends.
Upon ideal simulation, several programs like \texttt{less\_than\_qubit} still exhibit relatively low failure rates (e.g., below 0.25), indicating the presence of hard-to-trigger mutants and thus the need for systematic evaluation of testing techniques for such cases.

The comparison across backends reveals that the reported failure rates depend on backends even under the same test suite and oracle. 
In noiseless scenarios, DOSS has been shown to reduce both false positives and false negatives compared with common statistical oracles~\cite{li2016dynamic}, so its failure reports can provide a useful reference for program-induced faults before backend noise is introduced.
For programs like \texttt{parity\_qubit}, DOSS identifies similar failure rates across backends, suggesting that the oracle could tolerate certain noise effects.
However, noisy backends mostly shift failure rates upward, indicating that some additional reports may be induced by backend noise rather than program faults alone.

Furthermore, Table~\ref{tab: statistical_faults} compares test failure rates across backends using the Mann--Whitney U test and the Vargha--Delaney effect size $\hat{A}_{12}$~\cite{vargha2000critique}.
For each pairwise comparison, only 4 or 5 programs demonstrate negligible effects, while 21 or 22 exhibit large effects.
The comparison between Algiers and Brooklyn yields large effects for 21 programs, showing that reported failures are shaped by the specific backend model rather than by noise presence alone.
This result indicates that fake-backend choice is an experimental variable in fault-detection studies, rather than a peripheral execution detail.
\AnswerToRQ{3.2}{\BenchmarkName supports fault-detection experiments that reveal program differences and backend properties.
Some programs expose hard-to-trigger faults, making them valuable for evaluating QST techniques.
In the noisy scenarios, reported failure rates can differ significantly across fake backends, highlighting the need to consider backend-specific noise in test result interpretation.}

\section{Threats to Validity}\label{sec: threats}

\textbf{Internal validity.}
Internal validity may be affected by human involvement in program filtering, testability refactoring, and data annotation. 
To reduce subjectivity, we defined explicit guidelines and resolved disagreements through consensus.
Refactoring may introduce deviations from the original implementations, whose risk is mitigated by preserving source links, keeping refactoring moderate, documenting changes, and conducting quasi-blind unit tests. These checks support conformance to program specifications but do not prove full semantic equivalence to every original behavior.

\textbf{External validity.}
External validity is limited by three factors. First, \BenchmarkName focuses on Qiskit programs without intermediate measurements, which aligns with the choice of current QST research. Second, our empirical study uses classical simulation, as is dominant in existing QST studies~\cite{li2026methodological}. Experiments on physical hardware may differ due to time-varying noise and weaker reproducibility. Third, our fault-detection experiment adopts mutation operators rather than real-world bugs, which is common in QST~\cite{li2026methodological}. Also, unlike Defects4J~\cite{just2014defects4j} in CSE and Bugs4Q~\cite{zhao2021bugs4q,zhao2023bugs4q} for quantum software stacks with classical programs, \BenchmarkName does not include real-world bugs, since the development histories of benchmark programs are rarely available.

\textbf{Construct validity.}
Construct validity may be impacted by concepts like real-world provenance, scalability, and testability, where they are determined through observable and traceable evidence rather than formal definitions and quantified metrics.
We clarify that the notion of real-world provenance relies on claims in publications, documentation in source code, and inclusion in official SDKs, instead of industrial quantum applications.
The scalability relies on the availability of classical inputs, while the testability is manually validated through explicit specifications, unit tests, and test-case examples.

\textbf{Conclusion validity.}
\revise{Our empirical study offers an initial assessment, but cannot cover all programs, test suites, backends, and QST techniques. 
Thus, the results should be interpreted as evidence of feasibility and testing-relevant phenomena rather than definitive conclusions about all QST settings.
To support follow-up studies, we make the experimental settings and results explicit and reproducible.}

\section{Lessons Learned}\label{sec: discussions}

\textbf{Functionality-aware testing.}
Unlike low-level quantum programs, the programs in~\BenchmarkName exhibit clear program-level functionalities. This makes it possible for QST studies to design testing approaches that take intended functionality into account. For example, \texttt{draper\_adder} in Table~\ref{tab: output_category_table} implements integer addition for $x$ and $y$, where both inputs are encoded in computational basis states. Testing it with qubits initialized in entangled states falls outside its intended input domain and therefore does not provide meaningful evidence about functional correctness.
Our findings in RQ1 also highlight a class of quantum programs that produce dominant but inexact outcomes, for which functionally matched test oracles may be more appropriate than strict equivalence checks over the full output distribution. Although the output probability oracle~\cite{ali2021assessing} is the most commonly used oracle in prior studies~\cite{li2026methodological}, its validity may be largely limited to programs with probability-distribution outputs, and it may fail to correctly identify phase outputs~\cite{li2016dynamic}. These results suggest that test oracles should be selected according to the intended functionality and output characteristics of the target program.

\textbf{Investigation of fake backends.}
Compared with the ideal simulator used in most existing QST studies~\cite{li2026methodological}, fake backends introduce hardware-related properties into test execution. RQ3 shows clear differences in test cost between the ideal simulator and fake backends, even in noiseless settings, which calls for further investigation. Based on RQ2, we also found that benchmark programs can be structurally complex, and that their execution complexity strongly depends on classical input arguments. This is likely related to their scalability and modular design.
Under these conditions, testing large benchmark programs on noisy simulators may incur substantial overhead that classical hardware can hardly afford. 

\section{Conclusion and Future Work}\label{sec: conclusion}
In this paper, we systematically build \BenchmarkName, a benchmark infrastructure for QST empirical studies on scalable quantum programs.
\BenchmarkName provides refactored benchmark programs, program specifications, test case examples, and standardized interfaces to improve benchmark usability and support functional testing. Our empirical study shows that~\BenchmarkName covers diverse program functionalities and scales, and that many benchmark programs exhibit potential applicability and reusability. The study also highlights two issues that deserve more attention in future work: test case design should be aligned with program functionalities, and backend choice should be treated as part of the experimental setting.
{In the future, we plan to extend \BenchmarkName with more diverse quantum programs and SDKs, and also make improvements based on the state of research and industrial practice.}

\section*{Data Availability}

The infrastructure and empirical-study data for~\BenchmarkName are fully available at~\cite{qolumbina_infrastructure,qolumbina_empirical}, respectively. A webpage is further provided at~\cite{qolumbina_documentation_link} generated from the repository~\cite{qolumbina_documentation_repo}.

\bibliographystyle{plainnat}
\bibliography{components/ref}

\end{document}